\documentclass[lettersize,journal]{IEEEtran}
\IEEEoverridecommandlockouts

\usepackage{threeparttable}
\usepackage{cite}
\usepackage{amsmath,amssymb,amsfonts}
\usepackage{algorithmic}
\usepackage{graphicx}
\usepackage{textcomp}
\usepackage[dvipsnames, table]{xcolor}
\usepackage{paralist}
\usepackage[inline]{enumitem}
\usepackage{caption}
\usepackage{makecell}
\usepackage{tabu, booktabs}
\usepackage{float}
\usepackage{subcaption}
\usepackage{comment}
\usepackage{todonotes}
\usepackage[breaklinks,colorlinks,bookmarks=true]{hyperref}
\usepackage{csquotes}
\usepackage{soul}
\usepackage{pifont}
\usepackage{multirow}
\usepackage{tcolorbox}
\usepackage{tabularx}
\usepackage{array}
\usepackage{colortbl}
\usepackage{siunitx}
\usepackage{svg}

\definecolor{magma_darker}{HTML}{fdc38a}
\definecolor{magma_dark}{HTML}{e15666}
\definecolor{magma_light}{HTML}{82247f}
\definecolor{magma_lighter}{HTML}{1f0c43}

\definecolor[named]{xBlue}{HTML}{18647E}
\definecolor[named]{xOrange}{HTML}{FF9B00}
\definecolor[named]{xGray}{HTML}{808080}
\definecolor[named]{xGreen}{HTML}{60B950}
\definecolor[named]{xRed}{HTML}{A30B37}
\definecolor[named]{xDarkBlue}{cmyk}{1,0.58,0,0.21}

\definecolor{cadetblue}{rgb}{0.37, 0.62, 0.63} 
\definecolor{grayish}{rgb}{0.6, 0.6, 0.6} 
\definecolor{redorange}{rgb}{0.91, 0.41, 0.17} 
\definecolor{limegreen}{rgb}{0.5, 0.8, 0.2} 

\hypersetup{
    colorlinks=true,
    linkcolor=xOrange,
    filecolor=magenta,      
    urlcolor=xDarkBlue,
    citecolor=xBlue,
}

\usepackage[a4paper, total={184mm,239mm}]{geometry}
\def\BibTeX{{\rm B\kern-.05em{\sc i\kern-.025em b}\kern-.08em
    T\kern-.1667em\lower.7ex\hbox{E}\kern-.125emX}}
    
\ExplSyntaxOn
\DeclareExpandableDocumentCommand{\convertlen}{ O{cm} m }
 {
  \dim_to_decimal_in_unit:nn { #2 } { 1 #1 } cm
 }
\ExplSyntaxOff
\usepackage{gnuplottex}
\usepackage{tikz}
\usepackage{mathtools}

\usetikzlibrary{matrix,calc,positioning,arrows}

\newcommand{\cmark}{\textcolor{green!70!black}{\ding{51}}}
\newcommand{\xmark}{\textcolor{red}{\ding{55}}}

\newcommand*\circled[1]{\tikz[baseline=(char.base)]{\node[shape=circle,fill,inner sep=0.5pt] (char) {\textcolor{white}{#1}};}}

\begin{document}

\title{A Vertical Approach to Designing and Managing Sustainable Heterogeneous Edge Data Centers}

\author{
\IEEEauthorblockN{Aikaterini~Maria~Panteleaki, Varatheepan~Paramanayakam, Vasileios Pentsos, Andreas Karatzas,~\\ Spyros Tragoudas, Iraklis Anagnostopoulos}~\\~\\
\IEEEauthorblockA{School of Electrical, Computer and Biomedical Engineering, Southern Illinois University, Carbondale, IL, U.S.A.}
}

\maketitle
\begin{abstract}

The increasing demand for Artificial Intelligence (AI) computing poses significant environmental challenges, with both operational and embodied carbon emissions becoming major contributors. This paper presents a carbon-aware holistic methodology for designing and managing sustainable Edge Data Centers (EDCs), based on three design principles that challenge the state-of-the-art optimization paradigms. Our approach employs vertical integration across the architecture, system, and runtime layers, balances operational and embodied carbon emissions while considering EDC performance as a co-optimization objective, rather than a constraint. At the architecture level, we propose carbon-aware and approximate accelerator designs to reduce embodied carbon. At the system level, we enhance resource utilization and adapt to real-time carbon intensity variations to minimize operational emissions. Finally, at the runtime level, we develop dynamic scheduling frameworks that adjust execution, based on energy constraints and carbon intensity.
\end{abstract}


\setlist{nosep}

\section{Introduction}

The rapid rise of Artificial Intelligence (AI) has sharply increased the demand for computing, raising serious environmental concerns. Over the past few years, both AI hardware infrastructure and Machine Learning (ML) models have expanded significantly, intensifying the environmental footprint of AI systems across their entire life cycle, from manufacturing to deployment and eventually their disposal ~\cite{wu2022sustainable}. Large-scale AI workloads contribute significantly to carbon emissions through two distinct pathways; \textit{operational carbon} is emitted during the AI model deployment and depends on the energy consumption throughout device operation, while \textit{embodied carbon} is associated with the entire life cycle of the system, from hardware fabrication to distribution and end-of-life processing. While past sustainability efforts have primarily focused on reducing operational energy use, recent studies highlight the growing impact of embodied carbon emissions, especially in modern sophisticated hardware platforms/accelerators, required for efficient AI deployment ~\cite{wu2022sustainable, gupta2021chasing}. The notable contribution of both factors necessitates a holistic approach that addresses both operational and embodied carbon for sustainable AI design.

Edge Data Centers (EDCs), strategically positioned at the network edge near data sources, offer a promising path toward greener AI by enabling local processing, reducing transmission energy, and improving integration with renewable sources. Unlike traditional large-scale data centers, that employ large-scale compute and storage facilities, EDCs distribute their workload across smaller-scale infrastructure. EDCs typically incorporate performance and functional heterogeneity, combining iso-ISA cores (e.g., big.LITTLE), graphic processing units (GPUs), and other hardware accelerators (e.g., TPUs, NVDLA), to efficiently handle diverse AI workloads with optimal resource utilization.
Edge devices typically have a high embodied carbon footprint due to hardware manufacturing, while large cloud data centers mainly contribute through operational energy use \cite{gupta2022act}. EDCs share traits from both due to their significant embodied footprint and the increased energy required to execute AI workloads. Because of this, EDCs are special systems that require a comprehensive sustainability approach.

Addressing this challenge demands a coordinated strategy across the entire system stack, that targets both carbon dimensions, rather than optimizing for standard sustainability metrics, like energy consumption alone. We propose a unified, carbon-first approach that considers optimizations across three design layers, to address this unique combination of embodied and operational carbon emissions:
\begin{inparaenum}[(i)]
\item \emph{\textbf{Architecture layer}}, describing the hardware design of AI accelerators, including the compute part with the processing elements, the memory hierarchy with local and global buffers as well as their interconnects,
\item \emph{\textbf{System layer}}, that determines the resource management strategy during workload mapping across the heterogeneous compute platforms, including CPUs, GPUs and accelerators, and
\item \emph{\textbf{Runtime layer}}, which controls task scheduling and power management during execution.
\end{inparaenum} 
Unlike prior work, that treats these layers independently or prioritizes only one carbon aspect, our method enables cross-layer decisions to minimize the total carbon footprint. We demonstrate how
this integrated approach improves sustainability across diverse
edge computing scenarios.
\vspace{-5pt}
\section{Related Works}

\begin{table*}[htbp]
\centering
\renewcommand{\arraystretch}{1.4}
\vspace{-10pt}
\begin{threeparttable}
\caption{Summary of Existing Works on Carbon-aware Frameworks for Sustainable AI}
\label{tab:comparison}
\begin{tabular}{>
{\arraybackslash}m{4.0cm}>{\centering\arraybackslash}m{2.6cm}>{\centering\arraybackslash}m{1.6cm}>{\centering\arraybackslash}m{1.3cm}>{\centering\arraybackslash}m{1.0cm}>{\centering\arraybackslash}m{1.0cm}>{\centering\arraybackslash}m{2.5cm}}
\hline
\multirow{2}{*}{\textbf{Carbon-Aware Framework}} & 
\multicolumn{2}{c}{\textbf{Carbon Footprint Reduction}} & 
\multicolumn{3}{c}{\textbf{Optimization Level}} &
\multirow{2}{*}{\textbf{Performance}}\\
\cmidrule(lr){2-3} \cmidrule(lr){4-6}
& \textbf{Embodied} & \textbf{Operational} & \textbf{Architecture} & \textbf{System} & \textbf{Runtime} & \textbf{Consideration}  \\

\midrule
Offline Energy-Optimal LLM \cite{wilkins2024offline} & \xmark & Indirect & \xmark & \cmark & \cmark & Co-Opt  \\
Clover \cite{li2023clover} & \xmark & \cmark & \xmark & \cmark & \cmark & SLA \\
 GreenLLM \cite{shi2024greenllm} & Amortize, Reuse & \cmark & \xmark & \cmark & \xmark & SLO \\
 GreenScale \cite{kim2023greenscale} & Amortize, Reduce & \cmark & \xmark & \cmark & \cmark & Constraint \\
 EcoServe \cite{li2025ecoserve} & Amortize, 4R & \cmark & \xmark & \cmark & \cmark & SLO \& Co-opt \\
 CarbonEdge \cite{wu2025carbonedge} & \xmark & \cmark & \xmark & \cmark & \cmark & SLO \\
 CarbonCP \cite{ke2024carboncp} & \xmark & \cmark & \xmark & \cmark & \cmark & Co-Opt \\
 SkyBox \cite{sun2024exploring} & Amortize, Reduce & \cmark & \xmark & \cmark & \cmark & Co-Opt \\
 Carbon Explorer \cite{acun2023carbon} & Amortize, Reduce & \cmark & \xmark & \cmark & \cmark & SLO \\ 
 ACT \cite{gupta2022act} & \cmark & \cmark & \cmark & \xmark & \xmark & Constraint \& Co-Opt  \\
 CORDOBA \cite{elgamal2025cordoba} & \cmark & \cmark & \cmark & \xmark & \xmark & Co-Opt  \\
\midrule
\textbf{This paper} & \cmark & \cmark & \cmark & \cmark & \cmark & Co-Opt \\
\hline
\end{tabular}

\begin{tablenotes}
\footnotesize
\item Co-Opt = Co-Optimization;  SLA = Service Level Agreement; SLO = Service Level Objective;  4R = Reduce, Reuse, Recycle, Recover
\end{tablenotes}
\vspace{-10pt}

\end{threeparttable}
\end{table*}

As AI systems consume more energy and resources, researchers have focused heavily on finding ways to reduce their carbon impact. These efforts include a variety of strategies applied across all levels of the computing stack.

Targeting \emph{Architecture design layer}, \cite{gupta2022act} models robust embodied carbon estimation and introduces carbon-aware optimization metrics for sustainable design space exploration. \cite{elgamal2025cordoba} extends this by demonstrating a comprehensive framework that minimizes both embodied and operational carbon through architectural design strategies. However, these approaches focus solely on hardware level optimizations without considering system and runtime management.

In state-of-the-art works, \emph{System-level} and \emph{Runtime design layer} optimizations are usually coupled, as resource allocation and execution management are inherently synergistic processes. Several works focus on carbon-aware LLM serving through dynamic resource allocation \cite{li2023clover}, \cite{shi2024greenllm}, \cite{li2025ecoserve} and sophisticated workload-based energy models for heterogeneous platform selection \cite{wilkins2024offline}. At datacenter scale, holistic approaches target both embodied and operational carbon through design optimization \cite{acun2023carbon}, digital twin-based resource provisioning \cite{cao2025adaptive}, and spatio-temporal workload migration in distributed heterogeneous datacenters \cite{zhang2025carbon}. Renewable energy integration is explored through geo-distributed modular datacenter designs \cite{sun2024exploring}, showing how architectural flexibility along with clean energy significantly reduce environmental impact. For edge computing systems, carbon-aware design exploration frameworks address DNN scheduling based on workload characteristics and carbon intensity variations \cite{kim2023greenscale}, optimal workload placement across EDCs \cite{wu2025carbonedge}, and workload distribution between edge devices and servers, using conformal prediction \cite{ke2024carboncp}.

All aforementioned approaches indicate significant progress in sustainable computing. In Table \ref{tab:comparison}, we highlight how their area of focus differ. While state-of-the-art carbon-aware methodologies achieve operational emissions reduction through system-level and runtime optimizations, they usually decrease embodied carbon in a superficial way, employing amortization strategies. Most approaches follow the 4R sustainability principles (Reduce, Reuse, Recycle, Recover) to extend the useful life of hardware and spread the manufacturing carbon over time, rather than directly reducing emissions during the fabrication process. The most critical observation is that no existing work provides actual cross-layer optimization across architecture, system and runtime levels simultaneously. Additionally, while all works consider performance, many treat this metric as a constraint, often formulated as Service Level Objectives (SLOs) or Service Level Agreements (SLAs), rather than a co-optimization objective. This strategy limits the solution space, as it may prevent methods from discovering carbon-performance trade-offs, particularly important for delay-sensitive EDC applications. Our approach addresses these gaps through vertical integration that systematically reduces both carbon types via synergistic cross-layer optimizations, while treating performance as a first-order design objective along with sustainability metrics.
\vspace{-10pt}

\section{Methodology}

To reduce the environmental impact of AI computing, we propose a vertical integration approach that incorporates carbon-aware frameworks across all layers of EDC design and operation. Our strategy is built on three main design principles:

\begin{inparaenum}
\item[\circled{1}] First, we employ \textbf{\textit{vertical integration}} across architecture, system and runtime layers, enabling cross-layer information flow and synergistic decision-making.  This allows information to flow between layers and supports coordinated decision-making. As a result, we can achieve the best overall balance between carbon impact and performance, instead of just optimizing each design layer separately.

\item[\circled{2}] Second, \textbf{\textit{direct carbon reduction}} targets the embodied carbon during design and manufacturing stages, achieving greater footprint reduction compared to amortization strategies that only spread fabrication impact over longer time periods. 

\item[\circled{3}] Third, \textbf{\textit{performance co-optimization}} considers system efficiency as a first-order optimization metric, rather than simply a constraint that specifies the worst acceptable performance level. Our strategy expands the solution space and enables the discovery of carbon-performance efficient configurations that constraint-based approaches cannot achieve, making our framework suitable for delay-sensitive edge applications.
\end{inparaenum}

\begin{figure}
    \centering
    \resizebox{0.5\textwidth}{!}{\includegraphics[width=\linewidth, clip]{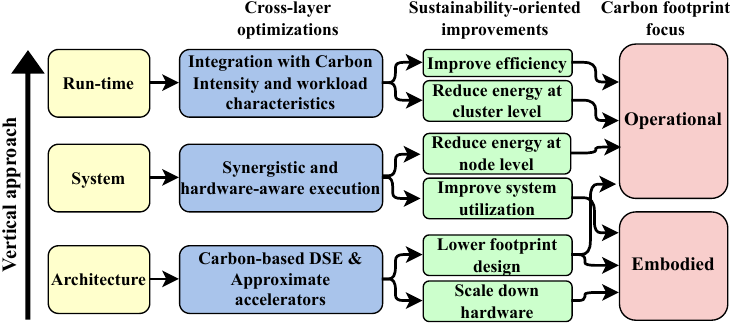}}
    \caption{Overview of the vertical optimizations along with the targeted sustainability parameters.}
    \vspace{-13pt}
    \label{fig:overview}
\end{figure}

Our methodology integrates these design principles through cross-layer optimizations targeting: (i) sustainable hardware accelerator design, (ii) improved system utilization to shorten carbon amortization, and (iii) runtime management based on real-time environmental conditions. 
These three layers exchange information and work synergistically, enabling holistic optimization of both embodied and operational footprint. 
Based on the workload properties, architecture-level decisions establish the proper hardware design, while system-level selections determine the optimal resource allocation. Finally, runtime adaptations enhance these layer-specific optimizations by dynamically adapting to environmental variations and minimize total carbon footprint, without impacting performance. Our integrated approach addresses the unique characteristics of EDCs, where both manufacturing and operational emissions contribute significantly to the overall environmental impact. Figure ~\ref{fig:overview} summarizes our methodology and its focus areas.

\vspace{-10pt}
\subsection{Architecture-Level Optimizations}

Our architecture-level framework introduces a unified carbon-aware methodology for designing efficient DNN accelerators under strict area and performance constraints. Unlike traditional hardware design pipelines that primarily consider energy efficiency and speed, we explicitly incorporate embodied carbon—the emissions from fabrication, materials, and packaging—into the design process as a first-class optimization objective. The embodied carbon footprint is formulated as:

\vspace{-5pt}
\begin{equation*}
    C_{\text{embodied}} = \sum C_{\text{die}} + C_{\text{packaging}} 
    \label{eq:emb}
\end{equation*}
where $\sum C_{\text{die}}$ are the carbon emissions from the manufacturing of all integrated dies and $C_{\text{packaging}}$ is associated with the employed packaging technique at the fab facility. Each die's manufacturing carbon emissions are calculated as:

\vspace{-10pt}
\begin{equation*}
    C_{\text{die}} = \text{CFPA} \times A_{\text{die}} + \text{CFPA}_{\text{Si}} \times A_{\text{wasted}}
    \label{eq:carbon_die}
\end{equation*}
\vspace{-10pt}

where, CFPA is the carbon footprint per manufactured unit area, $A_{\text{die}}$ is the area of the corresponding die component, and $\text{CFPA}_{\text{Si}}$, $A_{\text{wasted}}$ account for the silicon wastage on the wafer, due to shape mismatches. This formula reveals that chip area has the biggest impact on embodied carbon emissions, making area reduction a key strategy for sustainability.

Designing optimal DNN accelerators is generally a complex, multi-dimensional problem due to the vast design space of co-exploring hardware configurations and mapping strategies. The design space complexity is attributed to the inter-dependencies among hardware parameters, such as the Processing Elements (PEs) and memory hierarchy, and mapping techniques that determine data distribution across the architectural resources. To efficiently navigate this vast design space, we employ a genetic algorithm-based optimization framework that balances performance and sustainability ~\cite{panteleaki2024carbon}. Our fitness function is the Carbon Delay Product (CDP), a compound optimization metric that effectively illustrates the trade-offs between embodied carbon footprint and workload latency. The CDP metric enables the identification of Pareto-optimal accelerator configurations that deliver strong performance while significantly reducing embodied carbon footprint, rather than designs that minimize carbon at the expense of execution speed.

Our genetic algorithm employs a comprehensive chromosome encoding capturing all essential parameters for a complete DNN accelerator design. Each chromosome is structured as: 
\begin{equation*} 
\mathbf{C} = \{ P_x, P_y, B_{\text{local}}, B_{\text{global}}, M \} 
\end{equation*} 
where the hardware configuration, based on the renowned Eyeriss architecture \cite{chen2016eyeriss}, includes the PE array dimensions ($P_x, P_y$), the local buffer capacity (B$_{\text{local}}$) per PE as well as the global memory size (B$_{\text{global}}$). Additionally, each chromosome encodes the dataflow method ($M$) that determines how weights, input activations and output feature maps get distributed across the hardware platform. This unified encoding enables joint co-optimization for both architecture and mapping at the same exploration level, guiding the genetic algorithm to discover synergistic combinations that would be missed otherwise. 

To further enhance sustainability, we target the logic area of the chip by incorporating approximate computing into the accelerator design flow~\cite{10993191}. We extend our carbon-aware genetic algorithm framework to include approximate circuit selection, as an additional optimization dimension along with hardware and mapping. Leveraging the error resilience of DNNs, we selectively substitute exact arithmetic units with approximate multipliers within the Multiply-Accumulate (MAC) unit of each PE. Specifically, we target the multipliers, as they are more area intensive and dominate the silicon requirements compared to adder circuits, used for accumulation. Our methodology introduces a two-stage refinement process: first, we apply gate-level pruning and bit-width reduction to generate a diverse library of area-efficient approximate designs; second, we use a carbon-first optimization strategy to select configurations that maximize silicon area savings while keeping high performance, through the CDP optimization metric. Throughout this process, we systematically evaluate the inference accuracy impact of the different approximate multipliers using the method proposed in \cite{approxtrain-tcad}, to ensure that the selected configurations maintain acceptable accuracy for the target DNN model (typically $<$ 2\%). Experimental results across 7nm, 14nm, and 28nm technology nodes show that this approximation-based flow can reduce embodied carbon by up to 65\%, depending on the target model.

We extend this flow to support emerging design trends, such as 3D chiplet DNN accelerators. This architectural approach offers significant advantages in performance, energy efficiency and reduced 2D chip footprint, by enabling vertical integration of compute and memory resources \cite{yang2022three}. However, 3D integration introduces new sustainability challenges due to increased complexity in inter-chiplet bonding and advanced packaging requirements, mainly employing Through-Silicon-Vias (TSVs), which substantially increase the embodied carbon footprint compared to the 2D counterpart designs ~\cite{panteleaki2025carbon}. To address these challenges, we extend our genetic algorithm framework with 3D-aware carbon modeling that incorporates approximate multipliers. Our 3D accelerator design is based on the state-of-the-art accelerator model in \cite{wu202411}, where the bottom die hosts the PE array for computational logic, while the global SRAM memory is placed on the top die. We extend the embodied carbon footprint formulation to account for the additional 3D-specific emissions from bonding and TSV manufacturing. This extended framework jointly optimizes accelerator configurations through the CDP metric, combining architectural optimizations with approximate circuit selection, to maximize carbon gains while maintaining high performance. According to evaluations in 45nm, 14nm, and 7nm technology nodes, our carbon-efficient 3D accelerator designs achieve up to 30\% embodied carbon reduction with negligible inference accuracy loss, compared to baseline 3D configurations.

This carbon-aware hardware design forms the foundation of our vertical approach, creating a strongly sustainable architecture upon which system-level optimizations can build. The architecture layer provides crucial information to the system layer, including hardware properties and accuracy metrics, enabling scheduling decisions that account for both embodied and operational emissions to reduce the total carbon footprint.

\vspace{-4pt}
\subsection{System-Level Optimizations}

The sustainability of EDCs is increasingly dependent not only on the design of individual hardware accelerators but also on system-level management across heterogeneous architectures. Modern EDCs incorporate CPUs, GPUs and Deep Learning Accelerators (DLAs), that create a vast design space where only a very small fraction of possible multi-DNN mappings combine both performance and power efficiency ~\cite{karatzas2024mapformer}. Current practices often fail to fully utilize these resources, resulting in lower system utilization and prolonged amortization periods, directly impacting operational emissions ~\cite{gupta2021chasing}. The challenge becomes more complicated when we need to coordinate fine-grained layer-level partitioning across multiple concurrent DNN workloads, where traditional coarse-grained techniques fail to provide sustainable management strategies.

To address these challenges, our system-level optimizations target operational carbon footprint through energy-aware scheduling while co-optimizing performance as part of the vertical integration approach. The operational carbon footprint emitted during workload execution is formulated as:
\begin{equation*}
    C_{\text{operational}} = CI \times E 
    \label{eq:op}
\end{equation*}
where $CI$ is the carbon intensity of the electricity source and $E$ denotes the energy consumed by the system. This relationship reveals that minimizing operational carbon footprint is tightly connected to reducing energy consumption, analogous to how area reduction leads to lower embodied carbon footprint at the architecture level. However, unlike the static nature of embodied carbon, operational footprint varies dynamically according to energy consumption patterns and fluctuations in the $CI$, necessitating sophisticated resource management strategies.

For sustainable execution of AI workloads at the edge, we propose a system-level scheduling framework that dynamically coordinates the execution of multiple DNNs across heterogeneous embedded platforms. Our approach addresses the critical challenge of balancing high throughput and energy efficiency under constrained power budgets, with operational carbon reduction achieved through minimized energy consumption. This is achieved through a fine-grained scheduling mechanism that splits DNNs at the layer level and assigns each segment to the most suitable processing unit, CPU, GPU, or DLA, based on current system load and estimated energy cost~\cite{karatzas2024mapformer}.

A core innovation of our system layer is the integration of a transformer-based estimator that converts the mapping evaluation problem from regression to classification, predicting distributions over discrete classes for both throughput and power. This classification approach allows rapid evaluation of scheduling decisions without requiring expensive online profiling or retraining, enabling more accurate identification of high-performance solutions than regression-based cost models. Using this predictive capability, the scheduler explores the design space of possible allocations and operating frequencies, selecting configurations that maximize inferences-per-watt while ensuring that system constraints are met. By tightly coupling workload decomposition with hardware-aware predictions, our scheduler achieves up to 91\% improvement in energy efficiency compared to static mappings, effectively reducing the operational carbon footprint of DNN execution.

To further enhance sustainability in our vertical integration, we extend the scheduler to incorporate real-time forecasts of $CI$ from the energy grid~\cite{paramanayakam2025ecomap}. Our $CI$-aware framework dynamically adjusts the maximum power threshold based on 24-hour $CI$ predictions, where updates are triggered only when $CI$ changes exceed 10\% of the estimated range, to prevent oscillatory behavior. During periods with minimum $CI$, the system operates at maximum power threshold for optimal performance, while during periods with higher $CI$, the framework reduces the maximum power threshold and transitions to lower-power operation mode with reduced CPU, GPU and DLA frequencies. In addition, our framework supports adaptive model selection through mixed-quality model optimization that selects lighter DNN variants (e.g. ResNet-152 $\rightarrow$ ResNet-50), according to latency constraints and accuracy thresholds. This comprehensive approach ensures that operational decisions are not only performance- and power-aware but also carbon-aware. Evaluations show that the combined scheduling and $CI$-adaptation mechanisms reduce operational carbon emissions by 30\% and improve carbon-delay product by 25\%, all without compromising application-level latency.

Our system-level approach enables bidirectional information flow within the vertical integration framework. It firstly receives detailed hardware characteristics from the architecture layer to derive sustainable scheduling decisions through prediction models. It then forwards the optimized configurations to the runtime layer for validation and real-time adaptation. By bridging the gap between predicted and actual system behavior, this inter-layer coordination transforms the initial scheduling estimations into optimal sustainable strategies.

\vspace{-10pt}
\subsection{Runtime-Level Optimizations}

Our runtime-level optimizations complete the vertical integration framework by providing dynamic adjustments of system-layer scheduling decisions. Runtime management becomes critical as operational carbon emissions fluctuate based on workload characteristics, device configurations, and real-time energy grid variations, which cannot be fully captured by system level predictions. Traditional runtime systems apply static configurations, overlooking temporal variations in energy cost, workload intensity, and system-level constraints. Thus, intelligent runtime mechanisms that adaptively manage resources during task execution are essential to maximize energy efficiency and minimize operational emissions, especially under changing performance and environmental conditions.

The runtime-layer optimizations complete our vertical integration approach through system-level scheduling validation and real-time power management. To address the need for energy-aware task management, we first developed a dynamic scheduling framework for DNN inference workloads~\cite{pentsos2025energy}. Our runtime engine integrates a hierarchical scheduling mechanism that monitors latency, energy consumption, and concurrency during DNN inference. Using precomputed lookup tables of performance and energy profiles, the scheduler dynamically selects batch sizes, applies concurrent execution when beneficial, and adjusts GPU frequencies based on soft latency deadlines. This hierarchical optimization prioritizes energy-efficient execution without violating application deadlines. Experimental results show that dynamic batching alone can reduce energy consumption by up to 30\%, while intelligent concurrency and selective frequency scaling further reduce energy by an additional 15\%. Compared to static baseline execution, the proposed scheduler consistently meets deadlines while significantly lowering the total energy use, showing that runtime decisions can meaningfully improve sustainability outcomes.

We extend this methodology to support advanced AI workloads, such as large language models (LLMs) with real-time function calling~\cite{paramanayakam2025carboncall}. These models introduce additional runtime variability, as the computational load and latency requirements fluctuate based on input queries and tool invocation patterns, making runtime validation essential to ensure optimal sustainability. Our runtime controller implements a carbon-aware execution policy that dynamically selects among multiple quantized LLM variants depending on current power constraints and forecasted $CI$. When operational conditions are tight, such as during peak grid emissions, the system automatically falls back to lower-precision models and applies frequency scaling to maintain throughput while minimizing carbon impact. This integrated approach achieves up to 52\% reduction in operational carbon emissions, 30\% reduction in power, and 30\% improvement in execution time over conventional fixed-runtime execution strategies, demonstrating how runtime refinements elevate system-level scheduling from uncertain estimations to optimal sustainable models. These results validate our vertical integration approach by demonstrating how runtime adjustments build upon architecture and system-level optimizations to yield globally optimal sustainable solutions that layer-independent approaches would not achieve.
\vspace{-7pt}
\section{Evaluation}

We evaluate our vertical integration approach by assessing the frameworks of each layer. The results demonstrate the effectiveness of our carbon-aware methodologies across architecture, system and runtime levels for the key sustainability and performance metrics. In all cases, our frameworks achieve significant carbon gains for both operational and embodied emissions, while maintaining or even improving performance.

\subsubsection*{Architecture Level}
Figure \ref{fig:arch} presents the embodied carbon and latency trade-offs for our carbon-aware accelerator design framework. We evaluate three genetic algorithm-based approaches:
\begin{inparaenum}[(i)]
    \item Digamma \cite{kao2022digamma}, our baseline that co-optimizes hardware and mapping for minimal delay;
    \item GA-CDP, our proposed framework optimizing the Carbon Delay Product (CDP) metric;
    \item and GA-APPX-CDP, our framework extension that incorporates area-efficient approximate multipliers with up to 2\% inference accuracy drop.     
\end{inparaenum}
Results across VGG16 and ResNet50 models show that both our carbon-aware frameworks achieve significant embodied carbon reductions, compared to the baseline, with GA-APPX-CDP delivering the most prominent gains while simultaneously improving latency. This occurs because the genetic algorithm selects configurations with more processing elements that now have smaller area due to approximation, resulting in less embodied emissions and enhanced parallelism.

\subsubsection*{System Level}
In system level, we evaluate our enhanced carbon-aware scheduling framework against OmniBoost \cite{karatzas2023omniboost}, a throughput-optimized baseline and MapFormer \cite{karatzas2024mapformer}, a power-efficient multi-DNN manager. Our evaluation includes comprehensive testing across 50 widely used DNN models from different feature families, under varying carbon intensity and workload scenarios. Results in Figure \ref{fig:system} present the average performance across all tested configurations, demonstrating that our carbon-aware enhancements achieve 30\% less operational emissions and 25\% better CDP compared to the baseline, while maintaining competitive performance and power efficiency.

\subsubsection*{Runtime Level}
\begin{figure}[ht!]
    \centering
    \footnotesize
    \vspace{0.9em}
    \begin{subfigure}[b]{0.5\textwidth}
        \centering
        \includegraphics[width=\textwidth]{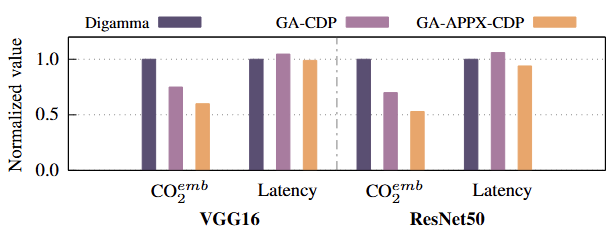}
        \vspace{-13pt}
        \caption{Architecture level}
        \vspace{1em}
        \label{fig:arch}
    \end{subfigure}
    \vspace{0.9em}
    \begin{subfigure}[b]{0.5\textwidth}
        \centering
        \includegraphics[width=\textwidth]{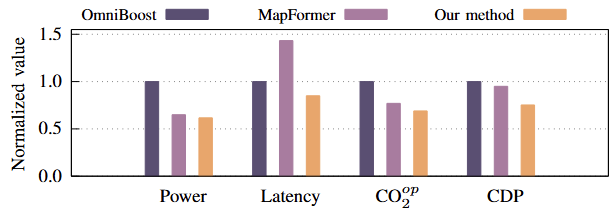}
        \vspace{-15pt}
        \caption{System level}
        \label{fig:system}
    \end{subfigure}
    \vspace{0.9em}
    \hspace{1mm}
    \begin{subfigure}[b]{0.49\textwidth}
        \centering
        \includegraphics[width=\textwidth]{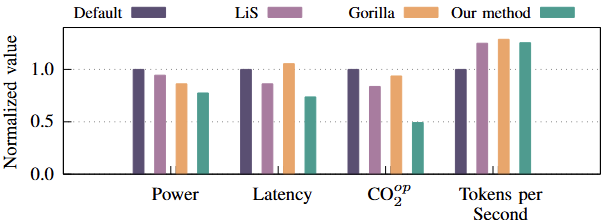}
        \vspace{-15pt}
        \caption{Runtime level}
        \label{fig:runtime}
    \end{subfigure}
    \vspace{-20pt}
    \caption{Overview of Architecture, System, and Runtime-Level Optimizations}
    \vspace{-15pt}
    \label{fig:combined}
\end{figure}

In runtime design level, we test our intelligent scheduling and carbon-aware dynamic function calling framework for LLMs against three state-of-the-art strategies:
\begin{inparaenum}[(i)]
    \item the Default case that represents the baseline LLM execution,
    \item the renowned API filtering method Gorilla \cite{patil2024gorilla},
    \item and the LLM-based tool recommendation framework Less-is-More \cite{paramanayakam2025less}.    
\end{inparaenum}
Our evaluation is employed across quantized LLMs with varying carbon intensity, while the average case results in Figure \ref{fig:runtime} demonstrate that our carbon-aware framework achieves 52\% reduction in operational emissions, 30\% lower power consumption and 30\% less execution time, compared to the Default baseline. The throughput metric of Tokens-per-Second (TPS) is maintained in acceptable levels through the dynamic scheduling of the quantized model variants.

\vspace{-10pt}
\section{Conclusion}

This work presents a holistic approach for sustainable AI computing at Edge Data Centers. Our contribution is based on three design principles, that challenge the dominant techniques of carbon-aware single layer optimizations. We employ vertical integration across architecture, system, and runtime layers, targeting direct embodied and operational carbon mitigation, while treating both performance and sustainability as first-order optimization objectives. Through synergistic optimizations across all layers of the computing stack, our holistic carbon-aware approach represents an essential paradigm for the future of sustainable AI infrastructure.

\section*{Acknowledgments}

This work is supported by grant NSF 2324854. Any opinions, findings, and conclusions or recommendations expressed in this material are those of the authors and do not necessarily reflect the views of the National Science Foundation.

\scriptsize

\end{document}